\documentclass[prl,twocolumn]{revtex4}

\usepackage{graphicx}
\usepackage{dcolumn}
\usepackage{amsmath}
\usepackage {epsfig}

\def\tp{$\tau$-phase }

\makeatother
\begin{document}
\title[Short Title]{High magnetic field-induced insulating
phase in an organic conductor}
\author{ J. S. Brooks, D. Graf, E. S. Choi, L. Balicas} \affiliation{National High Magnetic Field Laboratory
and Physics Department, Florida State University, Tallahassee, FL
32310 USA}
\author{K. Storr} \affiliation{Physics
Department, Florida A\&M University, Tallahassee, FL 32307 USA}
\author{C.H. Mielke} \affiliation{NHMFL/Los Alamos National
Laboratory, Los Alamos, NM 87545 USA}
\author{G.C. Papavassiliou}
\affiliation{Theoretical and Physical Chemistry Institute,
National Hellenic Research Foundation, Athens, 116-35, Greece}
\date{\today}

\begin{abstract}
We report electrical transport, skin depth, and magnetocaloric
measurements in the \tp series of organic conductors at very high
magnetic fields. Above 36 T these materials show a magnetic field
induced first order phase transition from a metallic to an
insulating state. The transition, which is a bulk thermodynamic
phenomenon, does not follow the conventional prescription for
field induced phase transitions in organic conductors.
\end{abstract}

\maketitle

    In low-dimensional metals, high magnetic fields often act to
reduce the effective dimensionality, thereby inducing a
metal-to-density wave state\cite{gorkov}. This is most evident in
quasi-one dimensional Bechgaard and related organic salts
\cite{ish} \cite{osada}. Even in the case of quasi-two dimensional
materials, the magnetic field plays an important role in
determining the ground state properties. \cite{harrison}
\cite{qualls} In such cases the interaction of the magnetic field
with the electronic structure is generally orbital in form, and
the magnetic field induced phases are therefore dependent on field
direction due to the low dimensional topology of the Fermi
surface. In marked contrast to conventional Fermi surface nesting
behavior, we present a description of a new type of field induced
metal-to-insulator transition in an organic metal.

Our main experimental result is the universal behavior of the
resistance of all isostructural members of the $\tau$-phase class
of organic conductors \cite{ter,pap95} in very high ($B
> 36 T$) magnetic fields. Here we find a first order
metal-to-insulator transition above a temperature-dependent and
hysteretic threshold field, $B_{th}$. An overview of the
field-temperature phase diagram of these materials is shown in
Fig. 1. What is remarkable about this transition is that $B_{th}$
appears to be only weakly dependent on magnetic field direction.
Hence the mechanism that drives the transition at $B_{th}$ may be
primarily isotropic, or Zeeman-like in origin, and we must look
for other physical characteristics of these materials which cause
this dramatic transition to the insulating state. The purpose of
this paper is to provide complementary electrical transport and
bulk thermodynamic information to describe this unconventional
magnetic field induced insulating state.

\begin{figure}[]
\epsfig{file=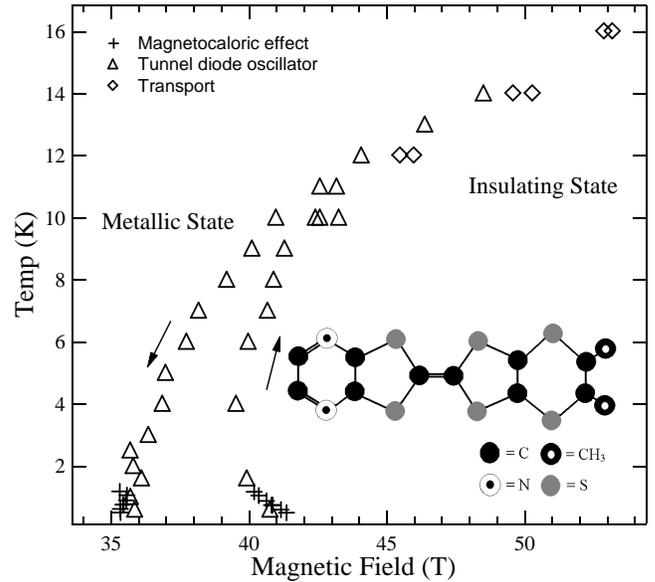, bb=72 322 508 712 clip=, width=\linewidth}
\caption{Schematic high field phase diagram of the \tp organic
conductors based on the P-DMEDT-TTF donor molecule with nitrogen
at the indicated sites (Fig. 1 inset). The threshold field
B$_{th}$ for the high field phase boundaries shown for
$\tau(r)$-AuBr$_{2}$ is based on magnetoresistance, ac skin-depth,
and magnetocaloric data. Arrows indicate the up-sweep and
down-sweep data.}\label{fig01}
\end{figure}

The complexity and structure of the $\tau$-phase unit cell is
unique amongst the general class of charge transfer salts (CTS)
\cite{will}.  Unlike conventional CTS materials with a charge
transfer ratio of 2:1, here it is 2:(1+y)(where $y \approx 0.75$)
and the anions occupy two different sites in the unit cell. A 2:1
ratio does exist in each conducting layer as (AuX$_2$) (where X =
Br and I) linear anion lies along the c-axis amongst a square
array of P-DMEDT-TTF donors. The (AuX$_2$)$_y$ anions are arranged
in the inter-layer ab-planes where their orientation alternates by
90$^o$ between layers. Since P-DMEDT-TTF is asymmetric (see Fig. 1
inset), the donor stacking involves alternating directions within
the each successive layer. Hence, due to the very low symmetry of
the donor and anion arrangement, four donor layers are necessary
to complete the unit cell\cite{pap95}, and the result is an
unusually large inter-planar dimension: $(a,b,c \approx
7.4,7.4,68~\AA)$.
$\tau$-(P-(\emph{S,S})-DMEDT-TTF)$_2$(AuBr$_2$)$_1$$_+$$_y$ and
$\tau$-(P-(\emph{S,S})-DMEDT-TTF)$_2$(AuI$_2$)$_1$$_+$$_y$
(hereafter referred to as $\tau$-AuBr$_2$ and $\tau$-AuI$_2$,
respectively) are completely analogous except for the replacement
of bromine for iodine in the anions. The prefix P refers to the
pyrazino (N-N) configuration of the P-DMEDT-TTF donor molecule.
Normally the salts of these structures are flat, but an uneven
hexagonal ring results when oxygen atoms (O-O) replace nitrogen to
produce
$\tau$-(EDO-(\emph{S,S})-DMEDT-TTF)$_2$(AuBr$_2$)$_1$$_+$$_y$
(referred to here as $\tau$-EDO). The direction of the twisting of
these rings is irregular making the EDO sample disordered. Further
disorder can be introduced into the $\tau$-phase system with a
racemic mixture of two isomers(\textit{R,R} \underbar{and}
\textit{S,S})~of the donor molecules \cite{pap01a,pap01b}. These
isomers differ only in the positions of the methyl groups with the
adjoining chiral centers. In general the ratio of the two isomers
is not 1:1, and this enhances the disorder over crystals
containing the pure (\textit{R,R} \underbar{or} \textit{S,S})
arrangements. The racemic system studied in the present work is
$\tau$-(P-(r)-DMEDT-TTF)$_2$(AuBr$_2$)$_1$$_+$$_y$ (hereafter
$\tau$(r)-AuBr$_2$). The compound
$\tau$-(P-(\emph{R,R})-DMEDT-TTF)$_2$(AuBr$_2$)$_1$$_+$$_y$ was
also investigated yielding very similar results to
$\tau$-AuBr$_2$.

    In this investigation, single crystals  of $\tau$-phase materials
(square plates of average size $1 \times 1 \times 0.2~mm^3$) were
grown by electrochemical methods. Resistance measurements were
four-terminal, inter-plane or in-plane measurements with current
values between 50 nA and 300 $\mu A$. Electrical contact was made
to the samples with 25 micron gold wires connected with carbon or
silver paint. A tunnel diode arrangement was used for the
skin-depth study, and a calorimeter platform in an evacuated
capsule was employed for the magnetocaloric study. Experiments
were carried out at the National High Magnetic Field Laboratory
(NHMFL) DC field facilities in Tallahassee, and at NHMFL-Los
Alamos with pulsed magnets.

\begin{figure}[]
\epsfig{file=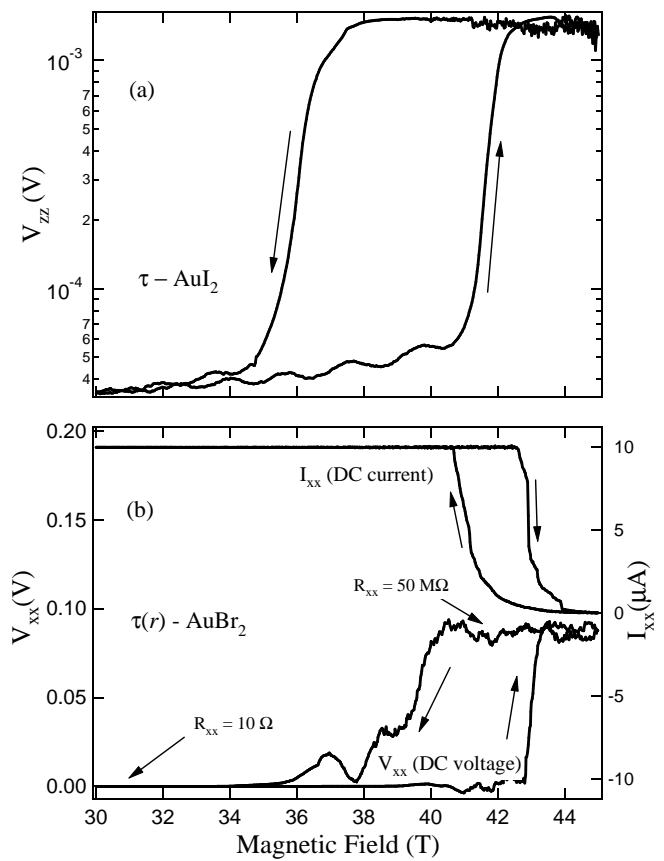, bb=70 115 509 694 clip=, width=\linewidth}
\caption{a) Inter-plane ($V_{zz}$) ac resistance data in dc
magnetic fields on $\tau$-AuI$_{2}$ at 0.5 K. Shubnikov-de Haas
oscillations are observed below the threshold field. b) In-plane
($V_{xx}$) dc resistance and current measurements for
$\tau$(r)-AuBr$_{2}$ in dc magnetic fields. The current (supplied
by a constant current source with a 1 volt compliance) vanishes
above 44 T due to the very high resistance of the sample.
Estimated resistance values for both the metallic and insulating
states are indicated. }\label{fig02}
\end{figure}

\begin{figure}[]
\epsfig{file=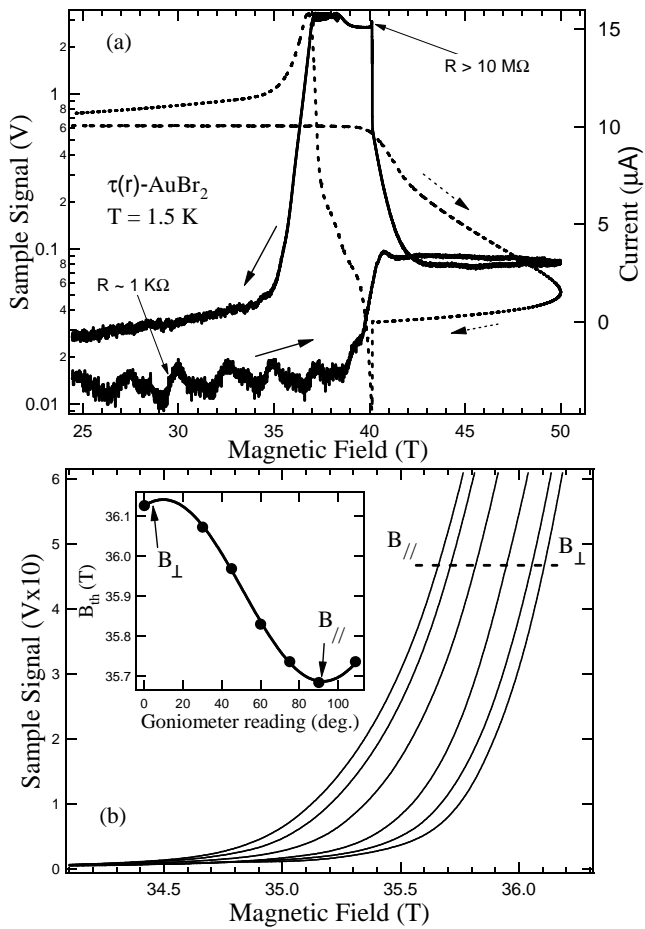, bb=84 75 529 718 clip=, width=\linewidth}
\caption{Inter-plane ($R_{zz}$) dc magnetoresistance studies of
the high field transition in pulsed fields. a) Simultaneous
measurement of sample voltage (solid line) and current (dashed
line) in $\tau(r)-AuBr_{2}$ at 1.5 K. Due to the fast time scale
of the measurement, the abrupt change in the sample resistance
produces dynamic changes in the current, although the current does
vanish in the insulating state. Estimated resistance values for
both the metallic and insulating states are indicated. b) Field
orientation dependence of the threshold field for the
insulator-to-metal (down sweep) transition. Inset: values for
$B_{th}$ vs. angle determined at a constant signal level (dashed
line). }\label{fig03}
\end{figure}

 The high field magnetoresistance of the
$\tau$-phase materials has been investigated in a dc hybrid magnet
up to 45 T ( Fig. 2), and in pulsed fields (6 ms rise-time) up to
60 T (Fig. 3). In all cases where the pyrazino (N-N) materials
were investigated, a transition between a metallic state and a
highly insulating state is observed at a threshold field $B_{th}$.
In Fig. 2a the inter-plane magnetoresistance for $\tau-AuI_{2}$,
is shown for low temperature ac transport data. For fields above
$B_{th}$, the inter-plane resistance $R_{zz}$ increases by orders
of magnitude, and in some cases becomes un-measurable by
conventional transport methods. $B_{th}$ is temperature dependent,
and moves to higher fields with increasing temperature. The
Shubnikov-de Haas (SdH) effect, which has been studied
previously\cite{sto} in $\tau-AuBr_2$, is observed below $B_{th}$
in this case. In Fig. 2b we show an in-plane dc measurement of
$V_{xx}$ on $\tau(r)-AuBr_{2}$ where the current is also
monitored. We find that the current vanishes in the high field
insulating phase.

    Inter-plane measurements of $\tau(r)-AuBr_{2}$ in pulsed
magnetic fields are shown in Fig. 3. In Fig. 3a we show the
simultaneous voltage and current signals. The transition, due to
the extremely abrupt increase (or decrease) of impedance of the
sample upon entering (or leaving) $B_{th}$, causes the current to
vanish (or reappear) on a short time scale comparable to the
response time of the coaxial leads. The strange appearance of the
voltage signal, which is characteristic of pulsed field
measurements for these materials, is primarily the result of the
current going to zero on the up-sweep. The resistance, as noted in
the figure, is obtained from the ratio $V_{zz}/I_{zz}$. In Fig. 3b
we show a detail of the insulator-to-metal signal for a systematic
variation of field directions from B//c to B//a-b plane. Both
hybrid and pulsed field results indicate that B$_{th}$ decreases
only slightly as the field is tilted into the a-b plane. In
contrast, for an orbital effect, the threshold field should
increase as $1/cos(\theta)$.

    To address the bulk-like nature of the transition, a
$\tau(r)-AuBr_{2}$ sample was investigated in a tunnel diode
configuration where the axis of the inductor coil and the c axis
of the sample were co-linear with the applied field. Hence the ac
field of the 43 MHz coil was perpendicular to the conducting
layers of the sample. In this configuration, a finite skin depth
of the ac field (typically about 1000 $\mu m$ for in-plane
conductivity in an organic conductor for this frequency range)
will reduce the resonant frequency of the TDO
oscillator.\cite{coffey} Measurements were carried out at both
high temperatures above the transition (18 K) to obtain a
background signal, and at low temperatures where the transition
was observed. The results are shown in Fig. 4a for the change in
frequency of the TDO vs. magnetic field. The results show that the
lower frequency observed in the metallic state increases,
essentially to an "empty coil" value, above $B_{th}$. From this we
deduce that the skin depth becomes larger than the coil, i.e., the
sample is a bulk insulator.

    We have further explored the bulk thermodynamic nature of the
transition in terms of the magnetocaloric effect, which has been
carried out under quasi-adiabatic conditions both in pulsed fields
and in the hybrid magnet. As shown in Fig. 4b for the hybrid
magnet, we observe an increase in the sample temperature as
$B_{th}$ is crossed for \underbar{both} directions. This result
eliminates the possibility that the transition is governed by a
Clausius-Clayperon type phase boundary with a latent heat, since
for decreasing field through $B_{th}$ the sample temperature
should decrease. Rather, the thermodynamics is governed by a
hysteresis-loop type of transition where the magnetic field does
work on the system, regardless of sweep direction. For
quasi-adiabatic conditions, a simple hysteresis model (based on
the functional form shown in Fig. 2a) yields a temperature signal
very similar to that shown in Fig. 4b, and magnetization
measurements are underway to further confirm this
picture\cite{brooks}.

\begin{figure}[]
\epsfig{file=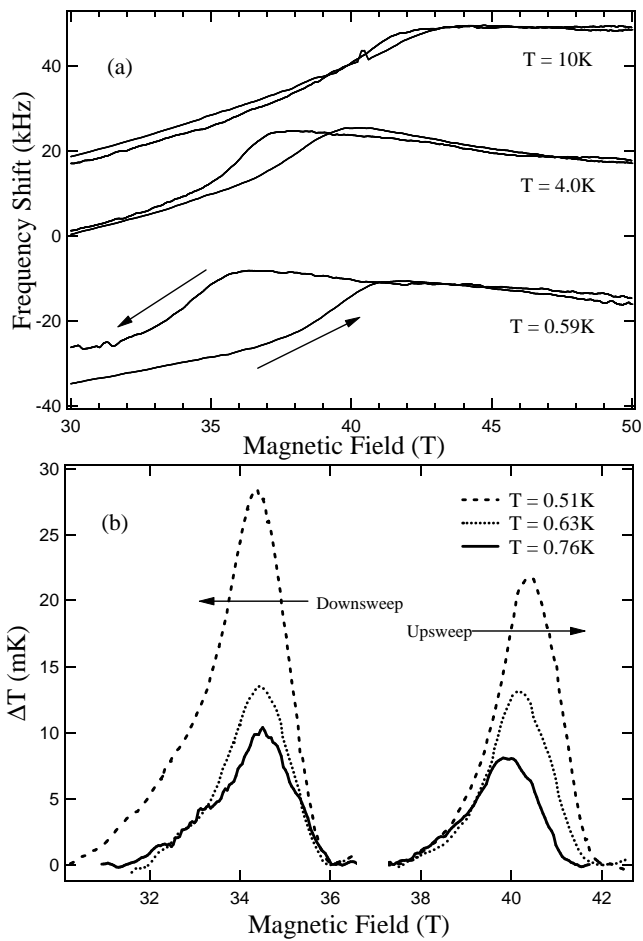, bb=75 70 517 712 clip=, width=\linewidth}
 \caption{Bulk properties of the high field transition. (See text for detailed discussion.)
 a) Response of a tunnel diode oscillator (TDO with background signal
removed) at 43 MHz to a $\tau(r)$-AuBr$_{2}$ sample in the
inductor coil in pulsed fields at different temperatures. The
sample orientation is $c//H_{dc}//h_{ac}$.  b) Magnetocaloric
effect (with background signal removed) in dc magnetic fields. A
temperature rise is observed for both increasing and decreasing
field at the hysteretic metal-insulating transition.
}\label{fig04}
\end{figure}

Clearly, magnetism is a possible source of the high-field
hysteretic behavior. Evidence for weak ferromagnetism has been
reported in the EDO class of \tp materials\cite{yosh1,kon1,kon2}.
(We note however that concurrent measurements \cite{brooks} in
this investigation have determined that the EDO class of \tp
materials, where O-O replaces the N-N site in the donor molecule,
show only a weak tendency towards an insulating state above 36 T.)
In addition, band magnetism has recently been suggested by Arita
et al.\cite{ari}, based on consideration of the very flat, narrow
($E_F \approx 8 meV$) bands that appear at the Fermi level from
the tight binding calculations. Although magnetization studies
have shown a small (0.001 $\mu_B$/formula unit) moment, no
hysteresis has been observed in low-field magnetization
data\cite{kon1}. Recent far infrared measurements indicate that
the methyl groups show disorder, and it has been suggested that
this may lead to electronic localization and magnetism at low
temperatures\cite{ole}.

We conclude that the high field metal-to-insulator transition in
the \tp materials has the following characteristics. It is most
pronounced in the N-N form of the donor molecule. By monitoring
the current in dc resistance measurements, we have determined that
the current vanishes above $B_{th}$ for both inter-plane and
in-plane transport. From skin-depth type investigations, we
conclude that above $B_{th}$ the sample is a bulk insulator, and
from magnetocaloric measurements, the transition is thermodynamic
in nature. Transport measurements in tilted magnetic fields
indicate only a weak dependence of $B_{th}$ on orbital effects,
which is completely contrary to behavior expected for conventional
field induced nesting-type behavior. The origin of the dependent
variable (i.e. the analog of $\chi H$ in the case of
ferromagnetism) in the hysteretic loop in the vicinity of $B_{th}$
remains elusive.  One possibility  would be a metamagnetic-type
transition\cite{stryjewski} where an anti-ferromagnetic spin
system is coupled to the lattice. This would provide the necessary
first order nature, required by the magnetocaloric data, for a
magnetic transition. The origin may be related to molecular bond
and/or conformation effects, given the N-N specificity of the
insulating transition. An investigation of the molecular structure
above 36 T, for instance by x-ray diffraction, could test for
conformational and lattice changes.

\begin{acknowledgments}
The FSU acknowleges support from NSF-DMR 99-71474. We would like
to thank P. Schlottmann, K. Murata, B. Ward, L. Gor'kov, E.
Dagotto, and R. McKenzie for valuable comments and suggestions.
The National High Magnetic Field Laboratory is supported by a
contractual agreement between the National Science Foundation and
the State of Florida.
\end{acknowledgments}
{99}
\end{document}